\documentclass[useAMS,usenatbib]{mn2e}
\usepackage{multicol}
\usepackage{epstopdf}
\usepackage{ulem} 
\usepackage{color} 

\def\lsim{\;\raise0.3ex\hbox{$<$\kern-0.75em\raise-1.1ex\hbox{$\sim$}}\;}
\def\gsim{\;\raise0.3ex\hbox{$>$\kern-0.75em\raise-1.1ex\hbox{$\sim$}}\;}
\def\lsimX{\raise0.3ex\hbox{$<$\kern-0.75em\raise-1.1ex\hbox{$\sim$}}\;}
\def\gsimX{\raise0.3ex\hbox{$>$\kern-0.75em\raise-1.1ex\hbox{$\sim$}}\;}

\definecolor{purple}{RGB}{200,100,255} 
\usepackage{ulem} 
\usepackage{color} 
\usepackage{graphicx}

\def\lsim{\;\raise0.3ex\hbox{$<$\kern-0.75em\raise-1.1ex\hbox{$\sim$}}\;}
\def\gsim{\;\raise0.3ex\hbox{$>$\kern-0.75em\raise-1.1ex\hbox{$\sim$}}\;}
\def\lsimX{\raise0.3ex\hbox{$<$\kern-0.75em\raise-1.1ex\hbox{$\sim$}}\;}
\def\gsimX{\raise0.3ex\hbox{$>$\kern-0.75em\raise-1.1ex\hbox{$\sim$}}\;}

\definecolor{purple}{RGB}{200,100,255} 

\newcommand{\Pev}{pevatron}
\newcommand{\Pevs}{pevatrons}
\def\flux4{\rm ~cm^{-2}~s^{-1}}

\newcommand{\Hess}{H.E.S.S.}

\newcommand{\Pmax}{p_\mathrm{max}}

\def\diff{\rm ~cm^2~s^{-1}}
\newcommand{\West}{\hbox{Wd1}}
\newcommand{\Msun}{\mbox{$M_{\odot}\;$}}
\newcommand{\tSNR}{t_\mathrm{SNR}}

\def\cmc{\rm ~cm^{-3}}

\def\cm2{\rm ~cm^{2}}
\def\ergs{\rm ~erg~s^{-1}}
\newcommand{\PP}{$p\!-\!p$}
\newcommand{\Tesc}{t_\mathrm{esc}}
\newcommand{\Tacc}{t_\mathrm{acc}}
\def\enf{\rm ~erg~cm^{-2}~s^{-1}~sr^{-1}}
\def\enf1{\rm ~erg~cm^{-2}~s^{-1}}
\def\enf2{\rm ~erg~cm^{-2}~s^{-1}~sr^{-1}}
\def\enfnu{\rm ~GeV~cm^{-2}~s^{-1}~sr^{-1}}
\def\IceC{{\sl IceCube South Pole Observatory}}
\newcommand{\gamray}{$\gamma$-ray}
\newcommand{\gamrays}{$\gamma$-rays}

\newcommand{\xx}[1]{\!\times\!10^{#1}}

\newcommand{\pcc}{cm$^{-3}$}
\newcommand{\kmps}{km s$^{-1}$}
\newcommand{\kmpsEq}{\mathrm{km\,s}^{-1}}

\newcommand{\TP}{test-particle}
\newcommand{\NL}{nonlinear}

\newcount\listnorom
\newcommand\listromanDE{\global\advance \listnorom by 1
{\lowercase\expandafter{(\romannumeral\listnorom)}\ }}

\newcount\listnumber
\newcommand\listDE{\global\advance \listnumber by 1
{\lowercase\expandafter{(\number\listnumber)}\ }}

\newcount\Inum
\newcount\IInum
\newcount\IIInum
\newcount\IVnum
\def\I{\global\multiply\IInum by 0 \global\multiply\IIInum by 0
            \global\multiply\IVnum by 0 \global\advance \Inum by 1
            {\the\Inum. }}
\def\II{\global\multiply\IIInum by 0\global\multiply\IVnum by 0
       \global\advance \IInum by 1 {\the\Inum.\the\IInum. }}
\def\III{\global\multiply\IVnum by 0\global\advance \IIInum by 1
            {\the\Inum.\the\IInum.\the\IIInum. }}
\def\IV{\global\advance \IVnum by 1
            {\the\IVnum. }}
\begin{document}

\title{Ultrahard spectra of PeV neutrinos from supernovae in compact star clusters}
\author[A.M.Bykov, D.C.Ellison, P.E.Gladilin and S.M.Osipov]{A.M.Bykov$^{1,2}$\thanks{E-mail:
byk@astro.ioffe.ru}, D.C.Ellison$^{3}$\thanks{E-mail:  don\_ellison@ncsu.edu}, P.E.Gladilin$^{1}$\thanks{E-mail: peter.gladilin@gmail.com} and S.M.Osipov$^{1}$\thanks{E-mail: osm2004@mail.ru}\\
$^{1}$Ioffe Institute of the Russian Academy of Sciences, Saint-Petersburg, Russia,\\
$^{2}$Saint-Petersburg State Polytechnical University, Saint-Petersburg, Russia,\\
$^{3}$North Carolina State University, Department of Physics, Raleigh, NC 27695-8202, USA}




\maketitle

\label{firstpage}

\date{\today}

\begin{abstract}
Starburst regions with multiple powerful winds of young massive
stars and supernova remnants (SNRs)
are favorable sites for high-energy
cosmic ray (CR) acceleration.
A supernova (SN) shock colliding with a fast wind from a compact cluster of young stars allows the
acceleration of protons to energies well above the standard limits of diffusive shock acceleration in an
isolated SN. The proton spectrum in such a wind-SN \Pev\ accelerator is hard with a large flux in the high-energy-end of the spectrum producing copious \gamrays\ and neutrinos in inelastic nuclear collisions.
%
%
We argue that SN shocks in the Westerlund 1 (\West) cluster in the Milky Way may  accelerate protons to $\gsim 40$\,PeV. Once accelerated, these CRs will diffuse into surrounding dense clouds and produce neutrinos with fluxes sufficient to explain a fraction of the events detected by IceCube from the inner Galaxy.
\end{abstract}

\section{Introduction}
Gamma-ray observations with both space-  and ground-based telescopes have shown that supernova remnants  are the main sources of CRs at least up to 100 TeV. Galactic sources of PeV CRs, however, are still to be identified \citep[see e.g.][]{blandford14,amato14} although it has been argued
that
type IIb supernovae, a subclass which comprise about 3\% of the observed core collapse SNe, may be able to produce CRs with energies beyond 100 PeV  \citep[e.g.,][]{ptu10}.
Cosmic neutrino observations are well suited for identifying  galactic \Pevs\
\citep[see e.g.][]{HH02, aharonian04,Becker08,pevatrons14}.
The {\IceC} (IceCube)
has detected  37 neutrino events   above the expected atmospheric neutrino background in a 988-day sample \citep[][]{Aartsen14}, with three PeV neutrinos, $1.041^ {+132}_{-144}$ PeV,  $1.141^{+143}_{-133}$ PeV and $2.004^{+236}_{-262}$ PeV, being the most energetic neutrino events in history.
While significant spatial or time clustering of the events has not yet been reported, a possible association of some events with galactic center sources was proposed \citep[e.g.,][]{Razzaque13}.

The {\sl ANTARES}  neutrino telescope \citep{ANTARES_GC_14}, using six years of data collected
near the galactic centre, reported 90\%
confidence level upper limits on the muon neutrino flux to be between 3.5 and 5.1 $\times 10^{-8}$ GeV $\flux4$, depending on the exact location of the source. They excluded  a single point source as the origin of 7 neutrinos observed by IceCube  in the vicinity of galactic centre. However, an extended  source  of a few degrees is not excluded.

Since the most likely high-energy neutrino
producing mechanisms are the inelastic $p-$nuclei and $p-\gamma$
collisions of protons, where the reaction kinematics result in the
energy of the neutrinos to be $\sim 0.05$ that of
the protons, the energy of the parent
protons should exceed $4 \cdot10^{16}$ eV to explain the IceCube observations. The \gamrays\ produced in these reactions have $\sim 0.1$ of the proton energy \citep[e.g.,][]{halzen13}.
Furthermore, the proton
accelerators must be very efficient to produce the high-energy neutrino flux of $\nu F_{\nu}\approx 10^{-8}\enfnu$ per flavor  in the
0.1-1\,PeV
range detected by IceCube.

Neutrinos from photo-meson $p-\gamma$
interactions in compact particle accelerators, like the cores of
active galactic nuclei (AGN) \citep[e.g.,][]{stecker13}
and \gamray\ bursts (GRBs) \citep[e.g.,][]{waxman97,MW01},
along with other models \citep[e.g.,][]{ahlers2013,Fox13,Kashiyamaea13,He2013,muraseea13,2014PhRvD..89h3004L,CRs_mergers_ApJ14,starburst_neutrinos14,neutrino_starburst14}, have been proposed to explain the origin of these first-ever IceCube neutrinos.

 It has been suggested by \citet{neronovea14} that  IceCube neutrinos and
Fermi/LAT \gamrays\ are both produced in interactions of CRs which have a hard spectrum with a power-law index harder than 2.4 and a cut-off
above $\sim 10$\,PeV. This assumes the CRs are interacting with the interstellar medium in the Norma arm and/or in the galactic bar.

The role of isolated SNRs in producing PeV CRs is uncertain.
The \gamray\ spectra from most shell-type galactic SNRs observed by  Cherenkov telescopes show a  spectral cutoff well below PeV energies  \citep[see e.g.][]{acero15}. While future {\sl  Cherenkov Telescope Array} observations may be needed to confirm these results
\citep[see e.g.][]{fa13}, a population of \Pevs\ is needed to explain the observed spectrum of galactic CRs at and above the spectral knee region.
Supernovae with transrelativistic  shocks \citep{budnik08,Chakraborti2011} and type IIb supernovae \citep{ptu10} were proposed  to accelerate CRs above PeV energies but  the statistics of these potential sources remain to be established.
Here, we present a model of a galactic \Pev\ which produces hard CR spectra with a high efficiency of conversion of SN kinetic energy into the highest energy CRs.
We further argue that this PeV CR source has the properties required to explain a number of the IceCube neutrinos detected from the direction of the inner Galaxy.
%


%
Core-collapse supernovae in associations of massive OB stars are certain to produce a  fraction of galactic CRs, as demonstrated by isotopic measurements  by the {\sl Advanced Composition Explorer}
\citep[see e.g.,][]{binns_SB07}.
Active star-forming regions may comprise  extended associations of massive young stars like Cyg OB2
\citep[see e.g.,][]{cygOB2_mnras14}, as  well as compact dense clusters of young massive stars like Westerlund 1 (\West). These extended and compact cluster types  are distinct in both spatial and temporal
scales \citep[e.g.,][]{clust_assoc_mnras11}. Both type of clusters of massive stars are expected to be efficient
PeV CR accelerators  in starburst and normal galaxies  \citep[see][for a review]{b14}.

The compact massive cluster \West, with an estimated age of
$3.5\!-\!5$\,Myr, contains more than 50 post main sequence stars
\citep[e.g.,][]{Clark_Wd1_AA05},  including at least 24  Wolf-Rayet (WR) stars of both flavours (representing about 8\% of the observed galactic population of WR stars) in about a parsec-scale core
\citep[e.g.,][]{Clark_Wd1_AA05,Wd1_WR_stars_06}.
These massive stars have strong individual winds with an estimated total cluster kinetic
power exceeding 10$^{38} \ergs$. In the small, dense cluster core these individual winds should combine and drive a fast cluster-scale wind by the mechanisms studied by \citet[][]{chev_clegg85} and  \citet[][]{stevens03}.
The magnetar CXOU  J1647-45, discovered  by \citet[][]{munoea06} using high resolution  {\sl Chandra} X-ray observations, has been associated with \West. This magnetar was likely produced about 10,000 years ago by a supernova with a progenitor star of mass $\geq$ 40 \Msun
\citep[][]{munoea06, mereghetti08} and  remains the only direct evidence of supernova activity in \West.

In our \Pev\ model, a SN blast wave collides with the termination shock of a strong wind generated by the collective action of many massive stars in a compact cluster.
Both shocks are assumed to propagate in a homogeneous upstream plasma.
We show that proton energies well above a PeV may be produced with a hard spectrum where the CR spectrum at PeV energies is most sensitive to the shock speeds and amplification of magnetic turbulence associated with CR driven instabilities. The model provides a  high efficiency of conversion of
SN shock--cluster wind ram pressure into PeV CRs which is needed to explain at least some of the IceCube neutrino events.
We don't consider a SN interacting with its own wind since individual stellar winds will be unimportant in the compact cluster environment where dozens of massive stars are located within a few parsecs and the  wind cavity is smoothed out on this scale.

We show  that diffusive shock acceleration (DSA)
in systems with colliding shock flows (CSFs)
can provide maximum particle energies, CR fluxes, and energy
conversion efficiencies well above those produced in an isolated SNR shock of the same velocity \citep[][]{MNRAS_BGO13}.
While CSFs are expected to occur in colliding stellar
winds, the most powerful events should happen when a supernova
shock impacts the extended fast wind of a nearby young
massive star cluster.

The important features of shock acceleration in \hbox{CSFs}
are:
(i) the production of
a piece-wise power-law particle distribution
with a very hard spectrum of confined particles
at the high-energy end just before a break, and
(ii) an increase in the maximum energy
of the accelerated particles, and the acceleration efficiency, compared to that obtained with DSA at an isolated SNR shock of the same speed.
These two properties imply that a substantial fraction of the flow ram
pressure is converted into relativistic particle pressure. The high-energy CRs, therefore, must modify the dynamics of the CSF system.
To model the spectra  of accelerated particles in CSFs, a
\NL, time-dependent model  was constructed in \citet{MNRAS_BGO13}. The maximum energy and absolute fluxes of the accelerated particles, both inside the CSFs, and those escaping the acceleration site, depend on the shock velocities, the number densities, and the magnetic fields  in the flows.
We show  below that sub-PeV and PeV neutrinos
from colliding shocks in galactic and extragalactic compact clusters
of young stars with reasonable parameters, such as those expected in the \West\ compact cluster, can explain a fraction of the IceCube neutrino events.
%
\begin{figure}  
\includegraphics[width=3.4in]{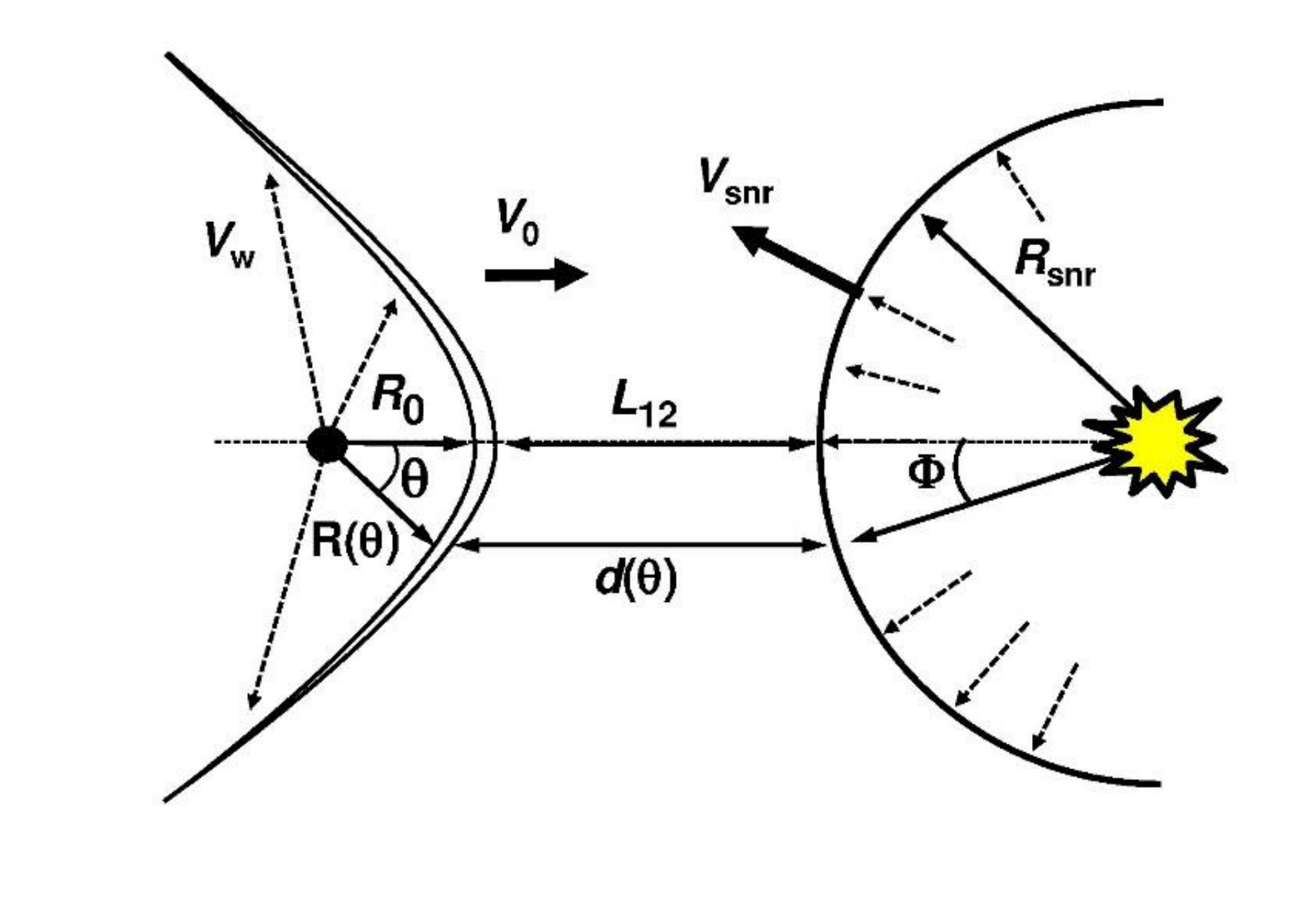}
\caption{Colliding flow geometry where the stellar wind (left) was
approximated from the analytic model of \citep{wilkin96}, while the SNR
shock was assumed to be spherical.
The star, or cluster center, approaches the SN explosion center with
speed $V_0$.
The spectrum of the Fermi accelerated CRs was derived taking into account
effects of the flow velocity projection on the curved shock surfaces.
}
\label{fig:3D}
\vspace{-1.\baselineskip}
\end{figure}
%

\section{Particle acceleration in colliding shock flows}
\label{sec:CSF}
In order to model the proton acceleration in a compact stellar cluster, we used the \NL, time-dependent model of CSFs presented in \citet{MNRAS_BGO13}. The main modification is that here we allow for different parameters for the wind termination shock and the SNR blast wave. We have also introduced an approximation to account for the  shape of the shock (as illustrated in Fig.~\ref{fig:3D}) using an analytic expression by \citet[][]{wilkin96}.
Other approximations, such as the  Bohm diffusion in the shock vicinity, and  a parametrization of the magnetic field amplification (MFA) due to CR-driven instabilities, are the same as in \citet{MNRAS_BGO13}. The reader is referred to that paper for full details. Our results for neutrino and \gamray\ production in \West,
given in Section~\ref{sec:WestI}, use our  \NL,
time-dependent model. However, to describe the general characteristics
of \hbox{CSFs},  we shall start  in this section with simple analytic estimates from a
linear model in a plain-parallel case \citep[e.g.,][]{bgo11}.

Considering a SNR expanding in a compact OB-association; at some expansion phase the distance $L$ between the SNR blast wave and a stellar wind shock is less  than the mean free path of the highest energy CRs in the SNR shock precursor.
At this point, the CR distribution function around the
two shocks, indicated by  $i = 1,2$, can be approximated as
\begin{equation}\label{SolveDC}
\begin{array}{l}
 f_i \left( {x,p,t} \right)= A p^{-3}\exp \left( {-\frac{u_i }{D_i }\left| x \right|}
\right)\times \\
\times  H\left( {p-p_0 } \right)H\left( {t- \Tacc } \right)
\ ,
\end{array}
\end{equation}
where the CR acceleration time is
\begin{equation}\label{TauAc}
\Tacc =\int\limits_{p_0 }^p {\frac{3}{\left( {u_1 +u_2 }
\right)}\left( {\frac{D_1 }{u_1 }+\frac{D_2 }{u_2 }} \right)}
\frac{dp}{p}.
\end{equation}
It is important to note that, besides only applying to high-energy particles with mean free paths larger than $L$, these equations are qualitatively different from those of \TP\ DSA in an isolated shock: the spectrum $f_i$ below the exponential
break  is harder and the acceleration time is shorter.

Our model assumes a high level of CR-driven magnetic instabilities and Bohm diffusion for CRs in the close vicinity of the shocks
\citep[see, e.g.,][]{bell04, schureea12, beov14}.
However, once high-energy particles obtain mean free paths on the order or larger than the distance between the shocks, $L$,
scattering will become much weaker.
A specific feature of the simulation is that
the highest energy CRs, with $\Pmax \geq  p \gsim p_{\star}$,
propagate with little scattering. Despite the weak scattering between the shocks, their momenta are still nearly
isotropic since they scatter for long periods in the extended regions downstream from the shocks. Here, $p_{\star}$ is the momentum such that the proton mean free path $\Lambda(p_{\star}) \gsim  L$.
Since the particle distributions are nearly isotropic
even for high-energy particles with  $\Lambda(p) \gsim  L$, the kinetic equation reduces to the so-called telegrapher  equation
\citep[e.g.,][]{earl74} and this allows a smooth transition between the diffusive and the
scatter-free propagation regimes.

The time-dependent nature of our simulation means that
protons escape the accelerator at different stages of the system evolution (i.e., as the SNR blast wave approaches the stellar wind termination shock) producing  pions and neutrinos with varying hardness and maximum energy
and these effects are included in our results.
Of course, the complex evolution of the source and some unknown details of the mass distribution in the outer ISM region (e.g., the presence of dense shells or clouds) will also influence the results. However, we believe the general properties of our simulation are robust.

Assuming Bohm diffusion with $D_1(p) =D_2(p) = D(p) = cR_g(p)/3$, due to CR-driven, amplified magnetic instabilities  in the CR accelerator, one obtains
\begin{equation}\label{TauAc1}
\Tacc \approx  \frac{c R_g(p)}{u_s\,u_w},
\end{equation}
where $u_1=u_s$ is the SNR shock velocity, $u_2=u_w$ is the
stellar wind speed, and $R_g(p)$ is the momentum dependent proton gyroradius.
Then, using the scaling $B \approx \sqrt{4\pi \eta_b \rho}\,u_1$ for the amplified magnetic field,
the acceleration time can be estimated as
$\Tacc \approx 2\,\cdot10^{10}\,
\epsilon_{\rm PeV}\, (\eta_b n)^{-0.5}\,u_{s3}^{-2}\,u_{w3}^{-1}$ s.
In the above expressions, $\rho=m_p n$, $\eta_b$ is the acceleration efficiency, i.e., the fraction of ram kinetic energy in the plasma flows converted into accelerated particles, the energy is in PeV, and the speeds are in units of $10^3$\,\kmps.

While the ejecta speed in the free expansion SNR phase can have a wide range of values, we use a mean ejecta speed of
$\sim 10^4 (M/\Msun)^{-1/2}E_{51}^{1/2}$\,\kmps\ and take the mean duration of the free expansion phase to be
$\sim 200 (M/\Msun)^{5/6} n^{-1/2} E_{51}^{-1/2}$\,yr.
In the specific case of a young SN shock propagating through the winds of massive stars in the compact cluster \West,
where $n \sim 0.6$\,\pcc\  \citep[see][]{munoea06}, and assuming $\eta_b \sim$ 0.1, we find $\Tacc \approx$ 400 yr
for a proton accelerated to
$\epsilon_{\rm PeV}\sim$ 40, when  $u_{s3} \sim 10$ and  $u_{w3}\sim 3$.
The high SN shock velocity $u_{s3} \sim 10$ is expected in the free expansion stage if the fast ejecta mass is about one solar mass.

Therefore, for SNRs with ages less than $\tSNR \sim 400$\,yr,
one can get $\Tacc < \tSNR$ for
40 PeV protons with standard parameters. In this case, at $\tSNR \sim 400$\,yr, the SNR radius is $\sim 3-4$\,pc.
Furthermore, the hard spectrum expected from CSFs puts most of the energy into the highest energy protons and one can estimate
the power in the highest energy neutrinos produced in
inelastic \PP--collisions by the decay of charged pions
($\pi^{\pm}\to e \nu_e\,\nu_\mu\,\bar\nu_\mu $) as
\begin{eqnarray}
&&L_{\nu} \approx 8\times 10^{33}\, \left(
\frac{f_{\nu}}{0.15}\right) \left(\frac{\eta_p}{0.1}\right)\,
\left(\frac{n}{1 \cmc}\right)^2 \times  \nonumber \\
&&
\left(\frac{S}{10^{38} \cm2}\right)\left(\frac{u_s}{5,000\,\kmpsEq
}\right)^3 \left(\frac{\tau_c}{10^{10}\, \rm{s} }\right)\,
\ergs ,
\label{eq:Lnu}
\end{eqnarray}
where $f_{\nu}$ is the fraction of energy in the inelastic
\PP--collisions which is deposited in the high-energy neutrinos,
$S$ is the cross section of  the colliding flows, and $\tau_c$ is the confinement time for protons in the emission region where the target density  is
$r_s \geq 4$ times the ambient density $n$ due to the shock
compression, $r_s$.

In Eq.~(\ref{eq:Lnu}), we take  the inelastic \PP\ collision
cross section to be $\sim 70$\,mb above a proton energy of 10 PeV.
Then the proton cooling time, $n\,t_{pp}$, can be
estimated as $n\,t_{pp} \approx 2.5 \cdot 10^{14}$ s assuming that two inelastic
collisions are needed to convert most of the proton energy into
secondaries. The fraction of the proton energy deposited into
neutrinos of all types, $f_{\nu} \sim$ 0.15, was derived using both the
 analytical parameterizations  for the inelastic \PP--collisions presented by \cite{kelner2006,cross_sect_PRD14} and
the {\sl  GEANT4} package simulations.
At the energies of photons and neutrinos considered in Figs.~\ref{fig:spectra_src}, \ref{fig:gam1m} and \ref{fig:NeuOnly}, the neutrino fluxes calculated with the \cite{kelner2006} or  \cite{cross_sect_PRD14} cross sections differ by less than 20\%.

We note that the observed CR energy density and the galactic SN statistics require that $\eta_p \gsim 10$\%  over the lifetimes of isolated SNRs to power galactic CRs.
However, the instantaneous efficiency may be much larger during early stages of the SNR evolution \citep{blasiAARv13}.
In the case of a SNR colliding with a wind,  we expect the efficiency to be  even higher than in a young isolated SNR and assume it can
reach $\eta_p \gsim$ 0.5 for the CSF stage lasting for a few hundred years \citep{MNRAS_BGO13}.
Furthermore, since CSFs produce very hard spectra when
$\Tacc \ll \Tesc$, i.e., $N(\gamma) \propto \gamma^{-1}$, most of the CR energy lies in the high-energy tail just below the upper break which occurs when $\Tacc$ is greater than the escape time, $\Tesc$.
The essential properties of CSF acceleration in young stellar clusters are high overall acceleration efficiency, spectra harder than $N(\gamma) \propto \gamma^{-2}$, and maximum proton energies $\gsim 40$\,PeV achieved in a few hundred years.

\begin{figure}  
\includegraphics[width=250 pt]{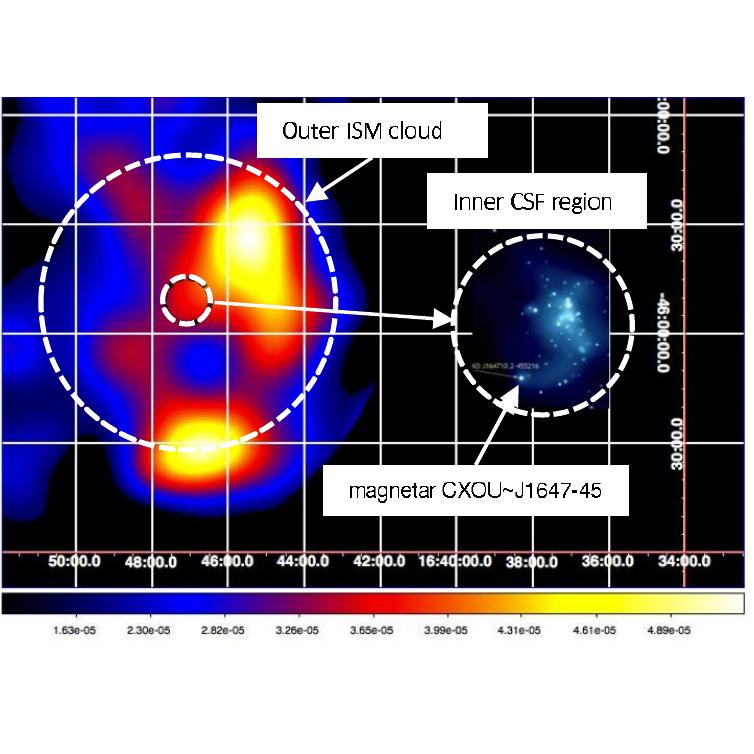}
\caption{The small dashed circles (not to scale) indicate the inner CR
acceleration region in the massive young cluster \West\. The inner CR
acceleration region has a radius of 3-4\,pc (shown in a Chandra image)
and  the blow-up of the inner region shows the position of the magnetar
CXOU  J1647-45 found by \citet[][]{munoea06} with a high angular
resolution Chandra observation of \West\. The outer  region is 30-40\,pc
centered around \West\  and indicated by the large dashed circle. This is
overlaid on a  H.E.S.S. map of TeV emission adapted from \citet{hessWd1}.
The angular resolution of IceCube is larger than the outer circle.
In Fig.~\ref{fig:Galactic} we show  the larger neutrino emitting ISM
volume of radius $\sim$ 140 pc around \West\. The volume is filled  over
$\sim 10^4$\,yr with  CRs  accelerated during a short CR acceleration
phase $\sim$ 400 years right after the supernova explosion  in \West\
which produced  the magnetar. }
\label{fig:West}
\vspace{-1.\baselineskip}
\end{figure}

\subsection{Colliding flows and 3D geometry effects}
\label{3D}
While an exact treatment of non-linear CSF with a significant backreaction from  accelerated CRs is unfeasible in 3D, we have introduced an approximation to account for effects from 3D geometry in our plane-parallel model.
Instead of taking the cluster wind termination shock and the SNR blast wave as plane, we account for aspects of the curved shock surfaces at positions parameterized by the angles $\theta$ or $\phi$ in Fig.~\ref{fig:3D}.
We still assume planar shocks for the DSA calculation but with varying projected velocities at distances $d(\theta)>L_{12}$ along the shock surfaces away from the symmetry axis.
Here, $L_{12}$ is the time-dependent minimum distance between the colliding shocks. For each position determined by $d(\theta)$  for $0<\theta<90^0$ we calculate the non-linear particle distribution using the projected speeds $V_w \cos\theta$ and $V_\mathrm{snr} \cos\Phi$ as parallel flow speeds. That is, we assume the wind and the SNR shocks are locally plane with converging parallel flows set by the projected speeds.

For this approximation, we
use the bow shock wind model of \citet[][]{wilkin96} for $R(\theta)$ and $V_w$, assume the SNR shock is spherical, and restrict our calculations to $3$ arc surfaces: $0-30^0$, $30-60^0$, $60-90^0$ which produce accelerated particle densities with the weights $0.84$, $0.15$ and $0.01$, respectively. Then the weighted CR distribution function is used to calculate the \gamray\ and neutrino emissivities  both in the accelerator (see \S\ref{sec:trap}) and  in the surrounding ISM from the CRs escaped the accelerator (see \S\ref{sec:esc}).

\begin{figure}                         
\includegraphics[width=300pt,trim=50 390 20 50]{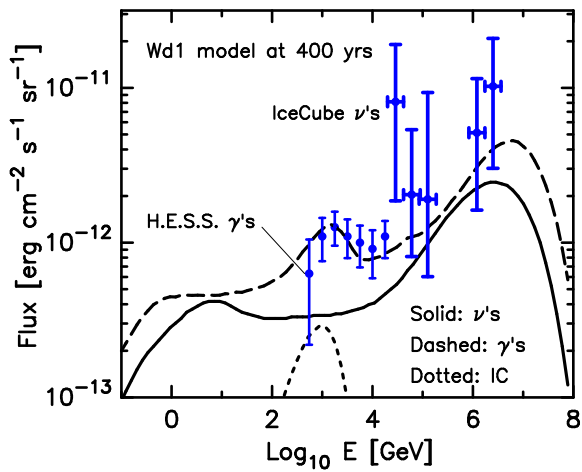}
\caption{Model predictions  of  \gamrays\  (solid curves) and neutrinos
(dashed curves)  from \PP-interactions calculated  in a CSF source of age
$400$\,yr. The dotted curve is the inverse Compton emission from primary
and secondary electrons accelerated directly in this source. The extreme
upward curvature in the neutrino spectrum above $\sim 10$\,TeV reflects
the transition from CR acceleration in the single SNR shock for
low-energy particles to the more efficient acceleration for high-energy
particles as they scatter back and forth between the SNR shock and the
cluster wind. The data points for the H.E.S.S. source, and the five  Ice
Cube events explained in
Fig.~\ref{fig:NeuOnly}, are presented to illustrate how they compare to
our model predictions when the source is $\sim 400$\, yr old and
point-like, i.e., \West\ about 10$^4$ years ago.
The simulated \gamray\ and neutrino emission at the present time from CRs
that escaped the accelerator in \West\ $\sim 10^4$\, yr ago are
summarized in
Figs.~\ref{fig:gam1m} and \ref{fig:NeuOnly}.}
\label{fig:spectra_src}
\vspace{-1.\baselineskip}
\end{figure}

\section{Neutrinos and Gamma rays from the Westerlund 1 Cluster}
\label{sec:WestI}
\subsection{Emission from Trapped CRs}
\label{sec:trap}
To explain the IceCube neutrinos, we combine the
\NL, time-dependent model of particle acceleration in CSFs with a propagation model applied to the \West\ compact cluster (Fig.~\ref{fig:West}).
The time-dependent simulations provide
the evolving energy spectra of CR protons and electrons as the SNR shock approaches the strong wind of a nearby early-type star.
For relativistic electrons/positrons we account for the energy
losses due to synchrotron and inverse Compton (IC) radiation.

The particle acceleration is combined with a propagation model relevant to \West.
As described in Section~\ref{sec:CSF}, high-energy
particles will obtain a hard spectrum when their mean free path is large enough so they scatter back and forth between the two converging shocks. Lower energy particles however, will be confined to, and accelerated by a single shock and obtain a softer spectrum. This phenomenon is illustrated in Fig.~3 of \citet{MNRAS_BGO13}.

It is important to note that even though the CSF acceleration is efficient and the CR population modifies the structure of the plasma flows, the spectrum of high-energy CRs scattering between the two shocks remains hard until they gain enough energy to escape from the system.
The shock modification can cause $\Tacc$ to increase, and the energy where the exponential
turnover in Eq.~(\ref{SolveDC}) starts to dominate drop, but below the turnover, the spectrum remains close to $N(\gamma) \propto \gamma^{-1}$.

Thus, the CSF system  will have
two spectral regimes for trapped CRs: a low-energy region ($\lsimX 1$ TeV) from particles accelerated in a single SNR shock (produced both before and after the
start of the two-shock acceleration period), and a
high-energy, hard spectrum region ($\gsimX 1$ TeV) from particles accelerated in the converging flows.
The transition between these two regimes of acceleration occurs
when the energy of a particle, $E_T$, is large enough so it can easily travel between the two shocks.

This transition, occurring at $\sim 20$\,TeV, is seen as a bend in the neutrino spectrum shown in
Fig.~\ref{fig:spectra_src}. This figure shows neutrinos (solid curve) and \gamrays\ (dashed curve) from \PP-decay from CR  protons that are still trapped near the SNR and stellar wind, along with IC from trapped CR electrons. This is the emission expected $\sim 400$\,yr after the SN explosion in a region of radius $\sim 3\!-\!4$\,pc. Note that the \gamray\ fluxes of the sources in Fig.~\ref{fig:spectra_src} are presented
 in ${\rm erg~cm^{-2}~s^{-1}~sr^{-1}}$ in order to be compared with the observed diffuse neutrino fluxes, while the fluxes of the sources
 in Fig.~\ref{fig:gam1m} are measured in ${\rm erg~cm^{-2}~s^{-1}}$ as usual.

In Fig.~\ref{fig:West} we have overlaid a schematic of this inner CSF region on a map of \West.
At later times, these ``trapped" CRs will escape the accelerator and diffuse beyond the inner CSF region into a much larger outer ISM region where they may encounter dense clouds
 producing neutrinos and \gamrays\ for an extended period of time, as we discuss in Section~\ref{sec:esc}.
For clarity, the regions are not drawn
to scale in Fig.~\ref{fig:West}.

To derive the gamma-emissivity of PeV CRs  we accounted for the Breit-Wheeler effect of pair production by energetic photons interacting with the interstellar radiation field as well as the extragalactic light background  \citep[see e.g.][]{aharonian04,Dwekea13}.
This interaction leads to a significant suppression of the
\gamray\ flux at the high-energy end of the spectrum for distant
($\geq10$ kpc) sources.

For the emission from \West, with an estimated distance of $\sim$ 4 kpc, this effect leads to a relatively small suppression of the \gamray\ flux at $1-10$ PeV, as can be seen in Figs.~\ref{fig:spectra_src} and
\ref{fig:gam1m}. In addition to the \gamrays\  produced by pion decay, we have also included \gamrays\ produced by the IC radiation of the secondary $e^{\pm}$ pairs produced {\it in situ} by the same \PP-interactions (see the IC curve in Fig.~\ref{fig:gam1m}). In this calculation, we accounted for the pair energy losses in a mean magnetic field of magnitude 10 $\mu$G in the extended cloud of number density 25\,$\cmc$ \citep[see e.g.][]{strongea14}.

\subsection{Emission from Escaping CRs}
\label{sec:esc}
The initial acceleration stage lasts a few hundred years  after the SN explosion producing high-energy CRs that escape the CSF system
and diffuse through the ambient ISM.
However, the estimated age of the supernova which produced the
magnetar CXOU~J1647-45 in \West\ is $\sim 10^4$\,yr
\citep[][]{NS_WesterlundI_muno06,mereghetti08}.
If the CSF acceleration occurred $\sim 10^4$\,yrs ago, these CRs will have produced pions over a much longer time span as the TeV-PeV particles diffuse away, fill a region of  about 140 pc radius, and interact with the ambient ISM (the outer ISM region indicated in Fig.~\ref{fig:West} is about 30 pc).

\begin{figure}                            
\includegraphics[width=300pt,trim=50 390 20 50]{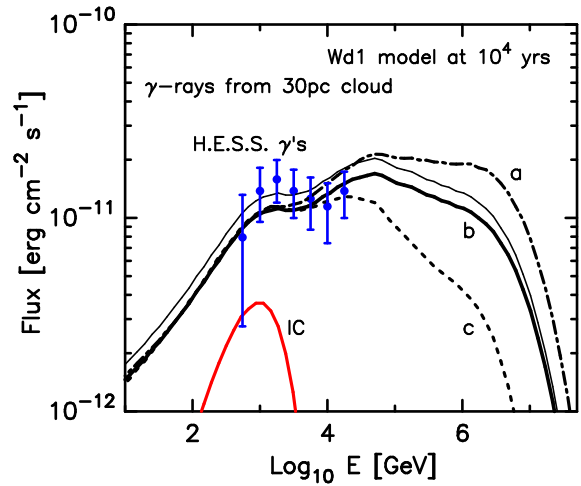}
\caption{Gamma-ray emission from inelastic \PP-interactions in the CSF
source at $\sim 10^4$\,yr after the SN explosion when CR protons produced
in the short-lived accelerator  have propagated into a nearby cloud
of $\sim 30$\,pc size. The magnetic field amplified by the CR-driven
instabilities in the vicinity of the fast shock in the CSF accelerator
were parameterized as  0.8 mG (c), 0.9 mG (b),
and 1 mG (a), all below 10\% of the ram pressure.
The IC curve is inverse Compton emission from the secondary electrons
produced by the inelastic \PP-interactions in the cloud. Only the
\gamrays\ from the H.E.S.S. field of view are included. The gas number
density of the nearby cloud is 25\,\pcc, except for the light-weight
solid curve where it is 30\,\pcc\ with $B=0.9$\,mG.}
\label{fig:gam1m}
\end{figure}

\begin{figure}               
\includegraphics[width=280pt,trim=50 140 0 250]{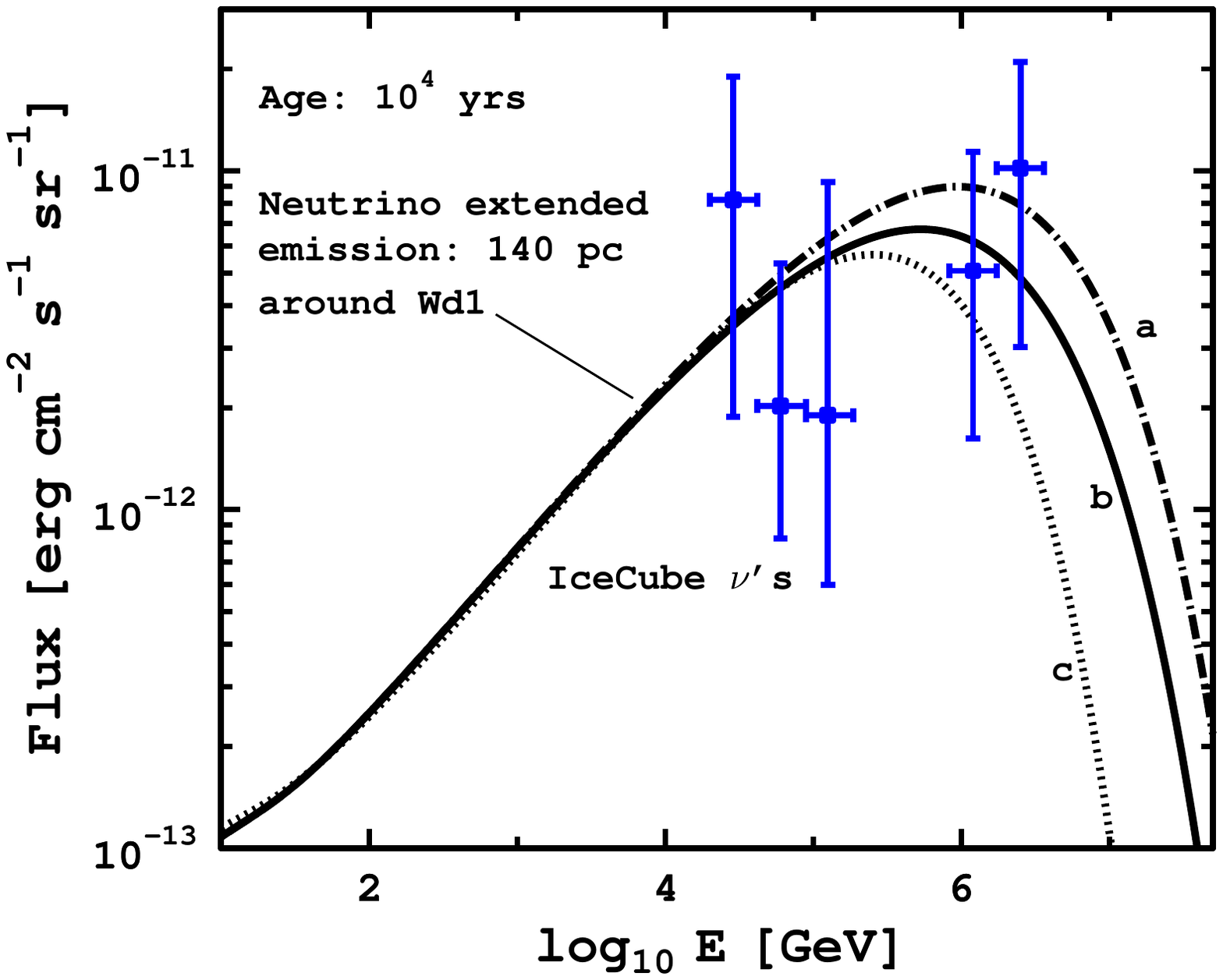}
\caption{Neutrino  emission from an extended ($\sim$ 140 pc radius)
source $\sim 10^4$\,yr after the SN explosion when CRs produced in the
short-lived accelerator  have propagated into the surrounding material.
The amplified magnetic fields are 0.8 mG (c), 0.9 mG (b),
and 1 mG (a). We note that the \gamrays\ in Figs.~\ref{fig:gam1m} and the
neutrinos here originate from different volumes: only the \gamrays\ from
the H.E.S.S. field of view are shown in Fig.~\ref{fig:gam1m}, while the
neutrinos are from a larger region of radius 140 pc. The  neutrino data
points (1$\sigma$ energy flux error bars) are a subset  from all 37
IceCube events \citep{Aartsen14}). These five events  are within
2-$\sigma$ contours  from \West\ based on Fig.~\ref{fig:Galactic}.
Two PeV events (14, ``Bert") and  35 (``Big Bird"), as well as three
sub-PeV (2, 15, 25)  events, are included in the subset. Note that the
sub-PeV event 25 has a very large position uncertainty in
Fig.~\ref{fig:Galactic} and have to be considered with some care.}
\label{fig:NeuOnly}
\vspace{-1.\baselineskip}
\end{figure}

In Figs.~\ref{fig:gam1m} and \ref{fig:NeuOnly}  we show simulated \gamray\ and neutrino spectra produced by the CRs that escaped from the accelerator and
diffused into the surrounding cloudy medium over $10^4$\,yr.
At this point the acceleration responsible for the emission in
Fig.~\ref{fig:spectra_src} has ceased long ago.

The diffusion model used to propagate the CRs has
three regions. Within the accelerator of 3-4 pc radius we assume a Bohm diffusion coefficient
$D_B = 3 \times10^{27} E_\mathrm{PeV} \diff$.
%
Outside of the dense cloud region we use an
ISM value $D_B = 3 \times10^{29} E^{0.33}_\mathrm{PeV} \diff$, which is consistent with the standard models of CR propagation in the Galaxy \citep[see e.g.][]{Strong2007}. Between these two regions
we assume a transition coefficient $D_\mathrm{tran} = D_B(R/3 pc)^2$,
where $R$ is the distance from the core. We assume a mean
cloud density $\sim 25$\,\pcc, cloud radius $\sim 30$\,pc, and have
set, at 1 PeV, $D_B = D_\mathrm{ISM}$ at $R = $ 30 pc. A cloud of radius 30 pc with mean density 25 $\cmc$ would have a mass
$\sim 10^5 M_{\odot}$. Outside the cloud we assumed an ISM density
of $\sim 1$\,\pcc. We note that the densities we assume for the clouds match available \gamray\ observations of
\West\ \citep{hessWd1,ohmea13}.

%
%
%

The IceCube data points in
Fig.~\ref{fig:NeuOnly} show the neutrino energy flux consistent with the position of \West\ considering the angular resolution of the instrument.
The CSF model applied to \West\ with reasonable parameters can explain a subset of the observed Ice Cube neutrinos.

\section{Neutrino PeV and sub-PeV events}
\begin{figure}
\includegraphics[width=280pt, trim=50 220 0 220]{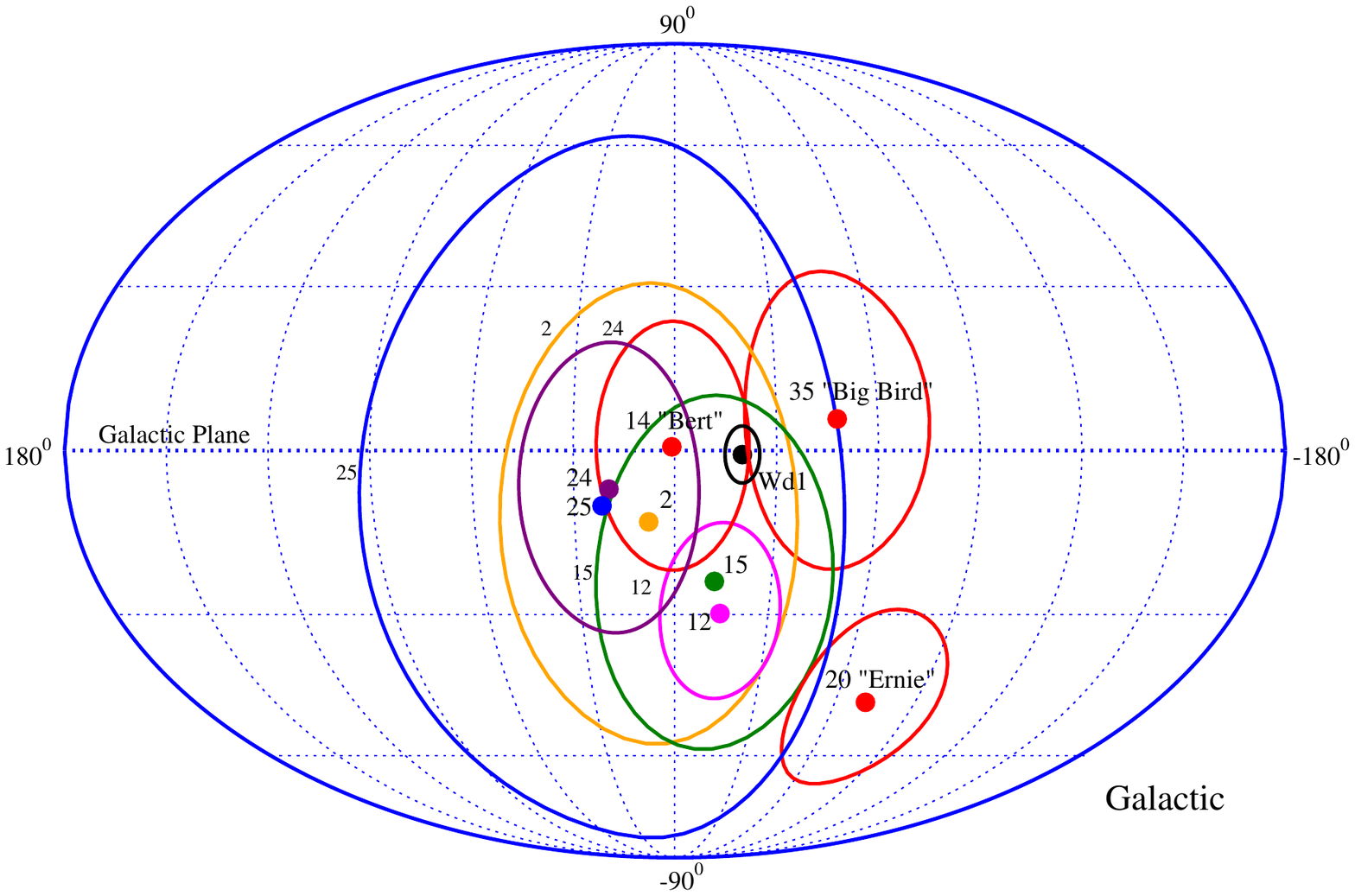}
\caption{Map showing 2-$\sigma$ contours for a subset of IceCube events
associated with the inner Galaxy as determined from 2D Gaussian
statistics and the median angular errors and positions given by
\citep{Aartsen14a}. Two PeV events (14, ``Bert") and  35 (``Big Bird") as
well as three sub-PeV (2, 15, 25)  events are within 2-$\sigma$ from
\West. }
\label{fig:Galactic}
\end{figure}

The median angular error values for the IceCube
neutrino events are given in the Supplementary Material for
\citet{Aartsen14}.
Using these median angular errors, and assuming Gaussian statistics
\citep[see][]{Aartsen14a}, we have produced a sky map
with $2\!-\!\sigma$ contours (corresponding to about $86\%$ confidence in 2D Gaussian statistics) for the neutrino events in
the vicinity of \West\ (see Fig.~\ref{fig:Galactic}).
As seen in this map, a source of neutrinos with a radius of $\sim 140$\,pc around \West\ (black circle) can be associated with five neutrino events, including two PeV events. These are $2$,
$14$ (PeV event ``Bert"),  $15$,  $25$ and $35$ (PeV event  ``Big Bird").

Based on this sky map, we compared our model neutrino spectra from
the \West\ source with the fluxes
in five IceCube energy bins corresponding to the five neutrinos within $2\!-\!\sigma$ of the source. This is shown in the Fig.~\ref{fig:NeuOnly}. A more precise comparison will require both more sophisticated models of the cloud
distribution within a few hundred parsecs of \West\ and a more accurate determination of the event
positions.

\section{Discussion}
Explaining the origin of the recently detected PeV neutrinos by IceCube is a fundamental challenge for models of particle acceleration. The observations imply a source that can produce substantial fluxes of protons with  energies considerably higher than those expected from isolated SNRs.
Since isolated SNRs remain the most likely source of CRs below a few PeV,
the neutrino source must produce high-energy protons without conflicting with the
observed properties of galactic CRs. The underlying protons producing the neutrinos in the CSF model described above  have a  hard spectrum, and are few in number, avoiding any conflict with low-energy CR population measurements. The
CSFs  may contribute to the high-energy end of the CR population produced by isolated SNRs and superbubbles \citep[see e.g.][]{binns_SB07} and, we believe the strong plasma flows in compact clusters of young stars, such as \West, contain the energy and specific properties needed to explain  a significant fraction of the IceCube neutrinos.

Compact clusters contain massive stars with strong winds and recent SN activity. It is inevitable that occasions will occur when a SN blast wave collides with the termination shock from the strong wind of a nearby massive star, or with an extended cluster wind from several massive stars. We have developed a model of the Fermi acceleration expected from such colliding shock flows and, using realistic parameters, obtained
simultaneous fits to the H.E.S.S. \gamray\ observations
(Fig.~\ref{fig:gam1m}) and the fraction of IceCube neutrinos expected from \West\ (Fig.~\ref{fig:NeuOnly}).

\subsection{Multi-wavelength signatures of CSF scenario}
A unique property of our CSF model is that the acceleration, while producing hard proton spectra to multi-PeV energies with high efficiencies, only lasts a small fraction of the SNR lifetime, just the time when the SNR is colliding with the nearby stellar wind.
%
%
Because of the hard spectra of accelerated CRs most of the energy is in the highest energy regime. In  Fig.~\ref{fig:spectra_src} we show the predictions for high-energy photon and neutrino emission of the source at the end of the acceleration stage (which was supposedly $\sim$ 10$^4$ years ago). The synchrotron radio emission then
is estimated to be  $\gsim$ 10\,Jy at 2.2 GHz, a value well above the current level.
This is consistent with our scenario where the brief CR acceleration stage ended $\sim 10^4$ years ago.

Indeed, the total radio flux from \West\ as measured with the
{\sl Australia Telescope Compact Array} (ATCA) interferometer
by \citet[][]{Wd1_radio_AA10} is 422, 461, 523, and 669 mJy at 8.6, 4.8, 2.2, and 1.4 GHz, respectively, and after subtracting  the radio emission from stellar sources, they derived diffuse emission fluxes of 307, 351, and 426 mJy at 8.6, 4.8, and 2.2 GHz.  Colliding winds in massive binary star systems were proposed by \citet[][]{eu93} to accelerate  relativistic particles and produce non-thermal radio and GeV regime \gamray\ emission. Some of the stellar radio sources detected in \West\ with the ATCA
by \citet[][]{Wd1_radio_AA10}
exhibited composite spectra of both non-thermal and thermal emission potentially indicating particle acceleration in colliding wind binaries.

\begin{figure}    
\includegraphics[width=280pt,trim=50 140 0 250]{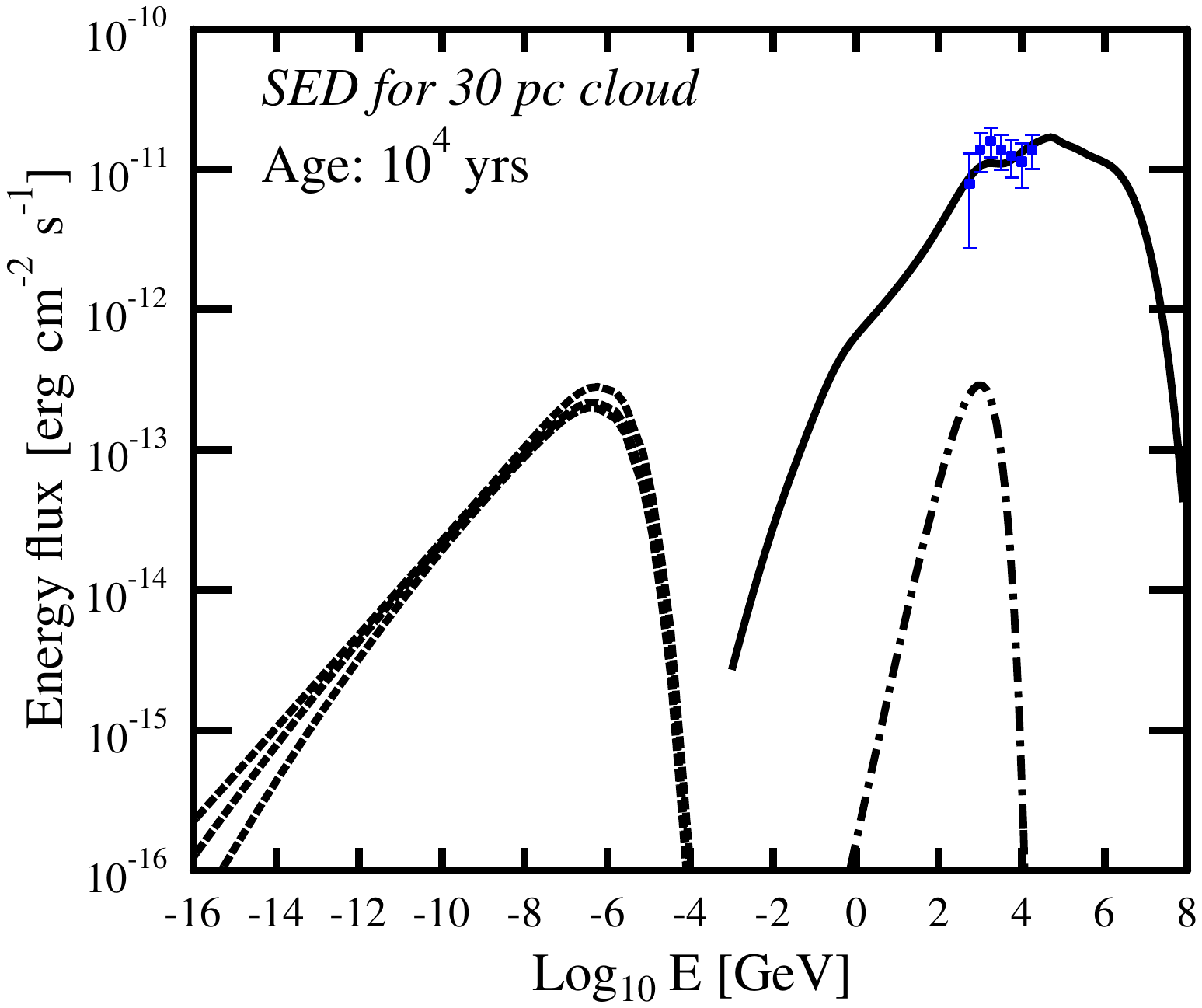}
\caption{The spectral energy distribution of the synchrotron (dashed
lines), inverse Compton (dot-dashed line) emission from secondary
electrons and positrons, as well as photons produced by pion-decay (solid
line) from the inelastic \PP-interactions in the nearby clouds which are
H.E.S.S. sources shown in  Fig.~\ref{fig:West}.  The CRs were accelerated
at \West\ and diffused into the clouds.  The cloud gas number density is
25\,\pcc\  with magnetic field  $B=10$\,$\mu$G. The \gamrays\ detected by
H.E.S.S. are indicated in the figure. The upper dashed synchrotron curve
is the result with no energy threshold for CR protons to penetrade into
the cloud. The lower synchrotron curves correspond to threshold values 10
GeV and 20 GeV respectively. The threshold values are low enough  to not
influence the pion-decay emission.}
\label{fig:SED}
\end{figure}

In contrast to the binary wind model, CRs in CSFs are generated in the violent environment of a SN blast wave colliding with a cluster wind. While the acceleration stage is brief, CRs will escape the source and interact with the nearby ISM clouds for long periods ($\gsimX 10^4$\,yr). 
%
During this time pions will be produced in inelastic \PP-interactions resulting in relativistic secondary electrons and positrons from $\pi^{\pm}$ decay and photons from $\pi^{0}$ decay.
%
The hard CR spectra from CSFs will result in prominent peaks in the spectral energy distribution produced by synchrotron, IC, and pion-decay emission, as shown in Fig.~\ref{fig:SED}.
For an extended cloud of size $\sim 30$\,pc and number density $\sim 25$\,\pcc, the peaks correspond to keV, TeV, and PeV energy bands.

For Fig.~\ref{fig:SED} we assumed the cloud magnetic field to be $B=10$\,$\mu$G consistent with Zeeman splitting measurements of a number of molecular clouds compiled recently by \citet{strongea14}.
Since the penetration of CR nuclei into the cloud may be reduced at low energies \citep[e.g.,][]{cv78,protheroe08}, we show in Fig.~\ref{fig:SED} spectra for the case with no proton penetration energy threshold (upper dashed curve) and for proton threshold energies
$E_{\ast} = 10$ and $20$ GeV.
The radio fluxes are sensitive to $E_{\ast}$ but the X-ray synchrotron fluxes corresponding to the peak of the synchrotron SED are not.

The cloud associated with the H.E.S.S. source in this model is a diffuse, low-surface-brightness ($\lsimX 0.1\,\mu$Jy arcsec$^{-2}$ at 1.4 GHz), flat-spectrum, synchrotron source of polarized radio emission with bright spots of brightness $\sim\,$ 1 mJy arcsec$^{-2}$ at 1.4 GHz, corresponding to local strong enhancements of the magnetic field in dense clumps.
The future {\sl Square Kilometre Array}, with a sensitivity of $\sim\,$ 1 $\mu$Jy/beam,  may allow detection of radio emission from clouds irradiated by CRs \citep[see e.g.][]{strongea14}.

The synchrotron peak  in the CSF model is at keV X-ray energies and a  source with a half-degree extension and total flux of $\sim 10^{-13}\, {\rm erg~cm^{-2}~s^{-1}}$ may be detectable with the future {\sl eROSITA} (extended ROentgen Survey with an Imaging Telescope Array)  instrument on the
{\sl Spectrum-Roentgen-Gamma} \citep[see e.g.][]{SRG} and {\sl ASTRO-H} \citep[see e.g.][] {astroH} satellites .
The search for synchrotron X-ray emission from the cold clouds located near these powerful CR sources may be conducted with the next generation of X-ray sky surveys.

In  Fig.~\ref{fig:gam1m} we show simulated spectra of \gamray\ emission from
CRs that escaped the accelerator and diffused into the
surrounding cloudy medium over $10^4$ yr. The models are
compared to H.E.S.S. data for the source associated with
 \West\  \citep[][]{hessWd1,ohmea13} and we have reduced the \gamray\ flux to correspond to the H.E.S.S. field of view at \West. At $10^4$  yr, the acceleration responsible for the emission has long ceased and the emission comes only from CRs accelerated in the source and propagating through the ISM .

We note that an analysis of 4.5 yr of {\sl Fermi-LAT} data by
\citet{ohmea13}  found extended emission  offset from \West\  by  about 1 degree.
This study concluded that acceleration of electrons in a pulsar wind nebula could provide a  natural explanation of the observed GeV emission. However, \citet{ohmea13} found that the pulsar wind nebula could not explain the TeV emission observed by  \Hess.
As seen in Fig.~\ref{fig:gam1m}, the CSF model can satisfactorily explain the TeV \gamray\ emission.

There is an apparent excess of neutrino events, including two of the three PeV neutrinos in the IceCube map presented in \citep[][]{Aartsen14}, within a radian
from \West. In Fig.~\ref{fig:Galactic} we show this set in a map with the positions of the events and 2-$\sigma$ contours as determined from 2D Gaussian statistics and the median angular errors of the IceCube telescope. Westerlund 1 is indicated with a contour (black circle) corresponding to a 140 pc radius region - a few degrees -
where neutrinos are produced in our CSF model as escaping CRs diffuse out from the compact accelerator for $\sim 10^4$\,yr.
The neutrino energy flux corresponding to the
solid curve in Fig.~\ref{fig:NeuOnly} is $\sim 3.7\xx{-8}$\,GeV\,$\flux4$
This is well below the 90\% confidence level upper limits imposed by  {\sl ANTARES} observations \citep[][]{ANTARES_GC_14} for a source of width $>0.5$ degrees  at the \West\  declination.

\subsection{CSFs in the starburst galaxies}
Another issue concerns
how hard spectrum PeV neutrino sources contribute to
starburst galaxy radiation. \citet{loeb_waxman06}  suggested that  CR interactions in starburst galaxies may efficiently produce high energy neutrinos and contribute cumulatively into the neutrino background. The population of star forming galaxies with AGNs and the starburst galaxies  peaked at a redshift $z \gsim$ 1, with a wide tail of the distribution revealed  by {\sl Herschel}
\citep[][]{Herschel_gal13} up to $z \sim$ 4.5.
These objects most likely contribute to both the  isotropic diffuse \gamray\ background  measured  by
{\sl Fermi-LAT} \citep[][]{Fermi_EGB15}  between 100 MeV and 820 GeV,  and the diffuse flux of high-energy neutrinos measured  by IceCube.

Assuming that a CR spectral index for all the starburst-like galaxies is 2.1--2.2  at the
high-energy part of the spectrum, \citet{starburst_neutrinos14}
were able to provide a reasonable fit to both the {\sl Fermi} and
IceCube data. Larger indices failed to explain the observed diffuse neutrino flux.
That CR spectra harder than in normal galaxies like the Milky Way were required, may reflect a different population of CR sources and/or different CR  propagation  in the starburst galaxies.
Superclusters of young massive stars are likely much more abundant in starburst galaxies compared to the  Milky Way since mergers and  interactions of galaxies result in abundant supercluster formation
\citep[see e.g.][]{Conti_book}.
The hard spectra and high efficiencies from CSFs make them an attractive way to produce CRs well beyond PeV in starburst galaxies where the collective contribution from many CSFs sources might extend the CR knee to higher energies compared to the Milky Way.

High resolution radio observations of the star forming galaxies M82,  Arp 220, NGC 253, M31, M33 and others provide information on the magnetic field structure and leptonic CRs  \citep{Tabatabaeiea13,m82_radio,pr14}, and
\gamray\ telescopes have observed some starburst galaxies up to $\sim 10$\,TeV
\citep[see][and the references therein]{acero_SFgalaxies_Sci09a,HESS_NGC253_ApJ12a,Lacki11,Fermi_SF_ApJ12a}. However, this is still well below PeV energies where CSF CRs are expected to be dominant and the  {\sl Cherenkov Telescope Array} \citep[][]{CTA11} would be needed to constrain the gamma-ray spectra in PeV energy regime.
More quantitative models of the CSF contribution to the diffuse \gamray\ and neutrino backgrounds will require better statistics of CSF SNe in starburst regions, accurate models of CR escape from the sources, as well as realistic models of CR propagation in starburst regions.

\section{Conclusions}
We have presented a colliding shock flow model for CR  production in compact stellar clusters that efficiently produces hard CR spectra and neutrinos.
Acceleration in colliding plasmas is a very efficient version of Fermi acceleration given the strong confinement of CRs in the converging flows. The mechanism is strongly  nonlinear and  time-dependent.
 We simulated the acceleration process in a simplified geometry.
Protons escape the accelerator with varying hardness and maximum energy as the
SN shock approaches the  wind. Furthermore, there is uncertainty in details of
the mass distribution in the complicated outer ISM region (e.g., dense
clouds) which will influence the results.
Nevertheless, we believe
our simulations include enough essential physics to estimate the neutrino and \gamray\ emission from the
galactic cluster Westerlund~1 and we show it is a
likely source for IceCube events detected from the inner galaxy.
Our \gamray\ predictions are consistent with H.E.S.S observations as well.

While the relatively large angular uncertainty in the arrival directions of PeV neutrinos precludes an exact identification, we believe some PeV IceCube events may result from $\geq$ 10 PeV CR protons accelerated in \West\ (see Fig.~\ref{fig:Galactic}).
This cluster is
a good candidate because it is one of the most massive
clusters in the local group of galaxies and has an observed
10$^4$ yr old magnetar, allowing enough time to spread PeV
CRs over a few hundred parsecs scale.

Future work that is critical for determining if a galactic supercluster can explain the apparent clump of 4-5
IceCube events includes developing a more accurate model of
multi-PeV CR diffusion on kpc scales.
This requires a careful treatment
of the matter distribution within a few degrees of \West\ and will result in a more accurate determination of the neutrino flux, as well as radio to \gamray\ emission, from CR interactions.

\section*{Acknowledgments}
We thank the anonymous referee for careful reading of our paper and
useful comments. A.M.B. thanks  Markus Ackermann for a useful discussion.
 A.M.B. and  D.C.E. wish to thank the International Space Science Institute in Bern where part of this  work was done. D.C.E. acknowledges support from NASA grant NNX11AE03G.

\bibliographystyle{mn2e}


\bibliography{Biblio_mn}
\end{document}